\documentclass{article}

\def\be{\begin{equation}}
\def\ee{\end{equation}}
\def\bea{\begin{eqnarray}}
\def\eea{\end{eqnarray}}

\def\d{\partial}

\def\tr{{\rm tr}}

\begin{document}

\title{Applications of the theory of evolution equations to general relativity}

\date{}

\author{Alan D. Rendall
\\ Max-Planck-Institut f\"ur Gravitationsphysik
\\ Am M\"uhlenberg 1, 14476 Golm, Germany
}

\maketitle

\begin{abstract}\noindent
The theory of evolution equations has been applied in various ways in general
relativity. Following some general considerations about this, some illustrative
examples of the use of ordinary differential equations in general relativity 
are presented. After this recent applications of Fuchsian equations are 
described, with particular attention to work on the structure of singularities 
of solutions of the Einstein equations coupled to a massless scalar field.
Next the relations between analytical and numerical studies of the Einstein
equations are discussed. Finally an attempt is made to identify fruitful
directions for future research within the analytic approach to the study of 
the Einstein equations.
\end{abstract}

\section{Introduction}\label{intro}

When general relativity is applied to model a physical system a connection
must be made on some level between solutions of the Einstein equations, in
general coupled to suitable matter fields, and the behaviour of the system 
concerned. As a consequence, a natural goal in general relativity is to 
obtain an overview of solutions of the Einstein equations which is as 
good as possible according to the criteria of generality, reliability and 
physical relevance. There are four main approaches to this goal which may 
be called algebraic, heuristic, numerical and analytical. In this article 
attention will be centred on the fourth of these but this subject will first 
be put in context by discussing the classification just indicated.

The approach referred to here as 'algebraic' is what is more commonly known
as the field of 'exact solutions'. A reasonable strategy when confronted 
with unfamiliar differential equations is to look for solutions in terms
of elementary functions. The basic work to be done consists of algebraic
manipulations, hence the terminology chosen. For sufficiently complicated
equations (e.g. the Einstein equations) the solutions obtained in this way 
are only a small subset of all solutions and this is why other approaches
are necessary.  In particular, it is not clear a priori how typical these
explicit solutions are. At the same time a solution which is given explicitly
in terms of an algebraic expression involving elementary functions may be 
less useful in understanding the nature of the dynamics than a qualitative 
analysis.

There are many heuristic approaches. One is to find a direct analogy 
between the equation being studied and the equations describing a
simple mechanical system. A well-known example is the analogy 
between the time evolution of a homogeneous and isotropic cosmological
model and the motion of a particle in a potential. Another procedure 
is to discard terms which are believed to be small in a particular 
situation and study the resulting simplified equation. One case frequently 
considered is that of small perturbations 
of an explicit solution. Here the stability of the original solution, and
thus its physical relevance, is being tested. A related but more general
approach is to try a particular ansatz and to see whether it is
formally consistent. The strength of heuristic approaches is that
they often furnish a wealth of detailed information. A problematic
aspect is that of their reliability. The success of a perturbative 
analysis in obtaining valid results often depends sensitively on the
intuition of the person carrying it out.

An attractive feature of numerical calculations of solutions of 
differential equations is that they are guaranteed to give results. These 
results may or may not be correct. Again we have a problem of reliability.
Numerical investigations can provide a quick way of checking a conjecture
and thus of helping to put a research project on the right track.  
In some situations it may turn out that there is no alternative to 
numerical calculations and then tests of the reliability of the results 
which are internal to the numerical calculations themselves are of 
particular importance. These may take the form of convergence tests
or of the comparison of two quite different numerical calculations 
which are expected to give the same answer. A good illustration of 
the latter procedure is the work of Gundlach\cite{gundlach95} which
provides one of the most convincing pieces of evidence that certain
aspects of critical phenomena in gravitational collapse are not just
numerical artefacts. It should be borne in mind that certain dynamical
features of solutions of partial (or even ordinary) differential 
equations might not be accesible to numerical calculation due to
limitations in resolution.

The application of the theory of partial differential equations is 
what will be referred to as the \lq analytical\rq\ approach in the
following. Frequently in textbooks phrases are encountered such as 
\lq no exact solutions are available and so we are forced to resort 
to approximate methods\rq. This sweeping judgement overlooks the value 
of qualitative information of the kind which can be obtained by the 
analytical approach. This article is concerned with explaining what the 
analytical approach has to offer and comparing it with the other 
approaches already mentioned. Where possible the fruitful interactions 
between the different ways of proceeding will be emphasized. The 
discussion is limited to the Einstein evolution equations and due to 
the limitations of what can be contained in one article the important 
topic of the constraints is hardly touched on here. It is the subject 
of the article of Isenberg in these proceedings.  

In the analytical study of the Einstein equations (or other partial
differential equations such as those of hydrodynamics) the concepts of 
existence theorems and well-posedness play important roles. Here only
those aspects of these ideas which are required in the sequel will be 
discussed. More detailed information can be found in \cite{friedrich00}. 
The properties which make up the definition of well-posedness of the 
initial value problem for an evolution equation are:
\begin{itemize}
\item The existence of a solution corresponding to any initial datum
appropriate to the problem at hand.
\item The uniqueness of the solution for a fixed initial datum.
\item The continuous dependence of the solutions on the data. (A small
change in the intial datum leads to a small change in the corresponding
solution.)
\end{itemize}
The realization that well-posedness is important goes back to Hadamard
in the first half of the twentieth century and it apparently took some
time for mathematicians to be convinced. Hadamard discusses the concept
in his book \cite{hadamard52}.

Ordinary differential equations with regular coefficients always have a
well-posed initial value problem. This is by no means true for partial
differential equations. It can be hard to prove well-posedness then.
(It should be noted in passing that if the coefficients of an ODE are
more or less singular then well-posedness cannot be taken for granted.)

A complication in studying well-posedness for partial differential
equations is that it depends on the function spaces chosen for the
data and the solutions. Related to this is the question of the sense
in which the mapping from data to solutions is supposed to be continuous.
This depends on the topology of the function spaces used. The type of
evolution equations of most importance in the study of the Einstein
equations are hyperbolic equations. The standard spaces used in this
context are the Sobolev spaces of $L^2$ type, i.e. the spaces of functions
which are square integrable together with their derivatives up to a
specified order $s$. This property defines the Sobolev space $H^s$.
In the limit of large $s$ we obtain the $C^\infty$ (smooth) functions.
Well-posedness in any of these spaces with $s$ sufficiently
large implies that the solutions are controlled pointwise by the
initial data. Another class of functions which often plays a role is 
that of  analytic ($C^\omega$) functions, i.e. those which can be expanded 
in a convergent power series about any point. It is relatively easy to 
obtain existence theorems when working in the class of analytic functions
using the theorem of Cauchy-Kowaleskaya. Unfortunately this does not
prove that the solution can be controlled pointwise by pointwise bounds 
on the data and finitely many of its derivatives. Another disadvantage
of analytic functions in this context is their unique continuation 
property which means that relativistic causality is obscured 
\cite{friedrich00}.
There is a class of spaces intermediate between analytic and smooth functions 
which are sometimes encountered. These are the Gevrey spaces. (See, for 
instance \cite{rodino97} for an account of spaces of this kind.) They 
share the disadvantage of the analytic functions that bounds on infinitely 
many derivatives of the data is required to estimate solutions. On the other
hand they do not have the unique continuation property.

The Einstein equations have the special properties that existence is
only obtained for initial data satisfying the constraint equations
and that uniqueness only holds up to diffeomorphism. The notion of 
continuous dependence of solutions on data must be correspondingly 
modified. For more details see \cite{friedrich00}.

The plan of the article is as follows. In section 2 various results
concerning ordinary differential equations are discussed. The third 
section explains the results of Lars Andersson and the author on 
constructing general spacetimes with singularities of a particular type. 
It also describes other applications of Fuchsian techniques in general 
relativity. Section 4 is concerned with the relations between analytical 
results and numerical relativity. The paper concludes with some ideas
about the future of the field.

\section{Ordinary differential equations}

Although the main concern of this paper is with applications of
partial differential equations, it should not be thought that
all problems in general relativity which require the analysis of
ordinary differential equations have been solved. Moreover,
examples concerning ordinary differential equations can serve
to illustrate issues which are also relevant for partial 
differential equations.

An interesting example of the interaction between numerical and
analytical methods is the case of the Bartnik-McKinnon solutions.
These are static spherically symmetric solutions of the 
Einstein-Yang-Mills equations. The problem of finding solutions
of this type can be formulated in terms of a system of ordinary
differential equations in a radial variable. Solutions which
are asymptotically flat were found numerically by Bartnik and McKinnon 
\cite{bartnik88}. These came as a surprise to many people. In this
way a numerical calculation led to a conceptual 
enrichment. Some time later an existence proof for these solutions was given;
the first result is in \cite{smoller91}. Unfortunately it does not appear 
that this existence proof led to new physical insights. It does not tell 
us more about why the solutions exist. A later alternative approach 
\cite{breitenlohner94} seems to be better in this respect. Another point 
of view, put forward by Corlette and Wald \cite{corlette01}, also offers 
an explanation, although it has not been made into an existence proof for 
the Bartnik-McKinnon solutions. (In related problems existence can be 
obtained in that way.)

For spatially homogeneous spacetimes the Einstein evolution equations
reduce to ordinary differential equations. The dynamics of the
solutions is very complicated, as exemplified by solutions of
Bianchi type IX, also known as the Mixmaster model. There is an
extensive literature where the Mixmaster model is studied by
heuristic and numerical means. This work led to a consensus on
what the main characteristics of the Mixmaster dynamics were but
until recently there were no rigorous proofs. This changed with
the work of Ringstr\"om \cite{ringstrom00a}, \cite{ringstrom00b} where 
the key qualitative statements, i.e. the presence of infinitely many 
oscillations in the approach to the singularity and the unboundedness of 
the curvature there in every solution except the Taub-NUT solution, were 
established. Thus finally, after thirty years, the analysis of the 
Mixmaster model came into the fold of rigorous mathematics. In the picture 
of singularities in general spacetimes due to Belinskii, Khalatnikov
and Lifshitz (BKL) \cite{bkl82} the Mixmaster solutions serve as the model
for the general case. For this reason the fact of having a 
mathematical understanding of this model acquires a special
significance.

There have also been other significant developments in the area
of the rigorous treatment of Bianchi models recently. Examples are the 
analysis of the expanding phase of models of type VII${}_0$ with perfect
fluid by Wainwright et. al. \cite{wainwright99}, \cite{nilsson00} and that of 
the expanding phase of Bianchi type VIII vacuum models by Ringstr\"om 
\cite{ringstrom01} The analytic investigation of spatially homogeneous 
spacetimes now appears to be reaching a certain maturity. At the same
time there are still many interesting open problems in this area. For
instance the spacetimes of Bianchi type VI${}_{-1/9}$, which are believed
to exhibit Mixmaster-like behaviour, have not yet been analysed rigorously.
(See \cite{wainwright97a} for more information on these models.) Also,
while the work of Ringstr\"om provides a lot of knowledge about untilted
Bianchi models with perfect fluids, models with tilted fluid have only 
been studied in special cases \cite{hewitt01}. A similar remark applies 
to Bianchi models with collisionless matter as source \cite{rendall99a},
\cite{rendall00c}.

Another interesting case where ordinary differential equations can be 
applied to general relativity is in the study of self-similar spherically
symmetric spacetimes. A good example is the proof by Christodoulou 
\cite{christodoulou94a} of the existence of spherically symmetric 
asymptotically flat
solutions of the Einstein equations coupled to a massless scalar field
which develop naked singularities. (His later proof \cite{christodoulou99a} 
that these solutions
are non-generic in the class of spherically symmetric solutions of course
uses partial differential equations.) For other models less is known
about self-similar solutions. The work of \cite{carr01} and references therein
provides a lot of information about this in the context of a convenient 
dynamical systems formulation but this has not yet been exploited to provide 
a full mathematical analysis of the problem.

\section{Fuchsian systems and spacetime singularities}

In \cite{andersson01} a class of solutions of the Einstein equations
coupled to a massless scalar field was constructed which are very
general and which have initial singularities which can be described
in detail. They provide the strongest mathematical support so far
for the heuristic analysis of spacetime singularities due to
Belinskii, Khalatnikov and Lifshitz. Near the singularity
in these spacetimes the evolution at different spatial points 
decouples, and the full evolution can be approximated by a
solution of a system of ordinary differential equations, the
velocity dominated system. The spatial coordinates play the role
of parameters. It is worth to remember that the BKL analysis is
an outgrowth of the earlier heuristic analysis of Lifshitz and
Khalatnikov \cite{lifshitz63} which gave the wrong answer (generic 
absence of singularities) and was shown to be in error by the rigorous 
mathematical arguments of the singularity theorems.

The statements of BKL concerned spacetimes depending on the maximum
number of free functions, and that is precisely the sense in which
the solutions whose existence is proved in \cite{andersson01} are
general. In fact the free functions in \cite{andersson01} are
required to be $C^\omega$, a restriction which does not seem
natural from a physical point of view. It would be desirable to
replace $C^\omega$ by $C^\infty$ and, as will be discussed below, 
there are indications that this should be possible. Counting
functions is a crude way of assessing the generality of a class
of solutions of a partial differential equation. A better 
characterization of generality is given in terms of open sets of
initial data. It should be shown that the class of solutions 
constructed include all solutions arising from a non-empty open set 
of initial data on a regular Cauchy surface. The possibilities of
doing this will be described below.

The method used in \cite{andersson01} is that of Fuchsian equations.
Fuchsian equations are a class of differential equations whose
coefficients have singularities of a certain type. They can be 
used to construct singular solutions of equations with regular
coefficients. Suppose that a regular system of equations, such
as the Einstein-scalar field system, is given. If an asymptotic
form for singular solutions of this system can be guessed an
ansatz can be made where the solution $u$ of the original system
is expressed in terms of an explicit singular function $u_0$ and
a remainder $u_1$ supposed to be regular. In favourable cases
writing the original equations in terms of $u_1$ leads to a
Fuchsian system, for which theorems on the existence of regular
solutions are available. In general, once the ansatz containing
some free functions has been introduced, the remainder $u_1$ is
determined uniquely.

In the case of the Einstein-scalar equations the ansatz for
the singular part is made in terms of a solution of the 
velocity dominated system. Velocity dominated quantities
are denoted by a left upper index zero. The starting point 
is a solution of the velocity-dominated constraints. This
consists of tensors ${}^0 g_{ab}$, ${}^0k^a{}_b$, ${}^0\phi$
and $\d_t {}^0\phi$ in three dimensions which satisfy the 
equations
\bea
&-{}^0 k_{ab}{}^0 k^{ab}+(\tr {}^0 k)^2=8\pi(\d_t {}^0\phi)^2  \\
&\nabla^a({}^0 k_{ab})-\nabla_a (\tr {}^0 k)=-8\pi\d_t {}^0\phi
\nabla_a{}^0\phi
\eea
In these equations raising and lowering of indices and covariant
derivatives correspond to the metric ${}^0 g_{ab}$. This system
of partial differential equations can be solved in analogy to
the full Einstein constraints by using the conformal method.
Now the velocity dominated evolution equations should be
solved for this initial data. This can be done essentially
explicitly, with the result that
\bea
&{}^0\phi(t,x)=A(x)\log t+B(x)           \\
&{}^0k^a{}_b(t,x)=K^a{}_b(x)t^{-1}         
\eea
after which the velocity dominated metric can be obtained by
solving the equation $\d_t {}^0 g_{ab}=-2{}^0 g_{ac}{}^0 k^c{}_b$.
This can be done using matrix exponentials.

The next step is to find a unique solution of the full 
Einstein-scalar field equations which approaches the given velocity
dominated solution as $t\to 0$. This can be achieved by applying the
Fuchsian theory, provided all the eigenvalues of $K$ are positive.
The resulting solution satisfies
\bea
&{}^0 g^{ac} g_{cb}=\delta^a_b+o(1) \\
&k^a{}_b={}^0 k^a{}_b+o(t^{-1})          \\
&\phi={}^0\phi+o(1)
\eea
Corresponding statements are obtained for space and time derivatives.
With this information the properties of the singularity can be
computed by straightforward algebra. Consider for example the 
curvature invariant $R_{\alpha\beta}R^{\alpha\beta}$. By the field
equations this is equal up to a constant factor to 
$(\nabla_\alpha\phi\nabla^\alpha\phi)^2$ The latter quantity is
equal to $A^4 t^{-4}$ up to an error of lower order in $t$. As a
consequence of the constraints $A$ can never vanish. Hence the
square of the Ricci tensor blows up uniformly as $t$ tends to zero
and the singularity is a curvature singularity.

Fuchsian techniques have been applied to a number of other problems
in general relativity and have the potential to be applied to many
more. In \cite{andersson01} results were obtained for the Einstein
equations coupled to a stiff fluid which are closely analogous to
those for a massless scalar field. In \cite{rendall00b} these results were
extended to the case of a scalar field with mass or potential. The
mass and potential terms become negligible near the singularity
and the velocity dominated system is identical to that obtained for
a massless scalar field. While the results just mentioned are the
only ones of this degree of generality obtained so far Fuchsian
techniques have also been applied to various situations with 
symmetry.

The theorem on Fuchsian systems applied in \cite{andersson01} was
proved in \cite{kichenassamy98}, where some background discussion
of Fuchsian methods can be found. The theorem there concerned only
the analytic case. It was applied in that paper to Gowdy spacetimes,
which are by now a standard testbed for applying new analytic 
techniques in general relativity. In \cite{rendall00a} the theorems
on the construction of Gowdy spacetimes with prescribed singularities
were generalized to the $C^\infty$ case. Some of the steps in the
argument did not depend on the details of the example being considered
and should be much more generally applicable. With luck it will be
possible to use them to handle the $C^\infty$ case for the 
Einstein-scalar system without symmetry. Some obstacles which
have to be overcome were listed in \cite{rendall00a}. Going beyond this
to the question of obtaining an open set of initial data without making
symmetry assumptions, no case involving the Einstein equations has been 
handled yet. There is, however, a treatment of this kind by 
Kichenassamy \cite{kichenassamy96} of the nonlinear wave equation 
$\nabla_\alpha\nabla^\alpha u=-e^u$ in Minkowski space. It was shown 
using the Nash-Moser theorem that there is an open set of initial data 
giving rise to singularities of a particular form. It is to be hoped that 
this technique can be extended to the Einstein equations. There are
already two results covering open sets of initial data for inhomogeneous 
spacetimes with particular symmetries \cite{chrusciel91}, \cite{rein96}.

In the case of spacetime singularities which show behaviour of 
Mixmaster type Fuchsian techniques cannot be expected to apply
and we have no method of constructing general spacetimes with
this type of singularity. In \cite{andersson01} the scalar field
was necessary and the approach would not work for a vacuum 
spacetime. It is worthwhile looking at what goes wrong. If
the scalar field is absent then it is like having a scalar field
with $A=0$. In that case it follows from the velocity dominated
Hamiltonian constraint that one of the eigenvalues of $K$ must
be negative and the assumptions of the theorem cannot be satisfied.
The significance of the sign condition on the eigenvalues of $K$
is that it allows the growth of the spatial curvature as $t$ tends
to zero to be estimated, thus showing that it eventually has a
negligible effect. It may be noted that it might be possible to handle
some negative eigenvalues and hence some vacuum solutions in higher
dimensions (spacetime dimension $\ge 11$) since the algebra works
out differently \cite{demaret85}.

There has recently been significant progress in understanding Mixmaster
behaviour in inhomogeneous spacetimes by heuristic and numerical 
techniques. See for instance \cite{weaver98}, \cite{berger98},
\cite{berger01}. Putting the results of this work on a rigorous
mathematical footing is an fascinating challenge for the analytical
approach to general relativity.   

The discussion in this section up to now has been from the point
of view of the BKL picture of spacetime singularities. In the
past decade, starting with \cite{poisson90} it has been suggested 
that the singularities of the kind studied by BKL, which are in some 
sense spacelike, are not sufficient to describe the singularities
inside black holes. Instead another type of singularities, called
weak null singularities, are supposed to play an important role.
Until recently the evidence for this was essentially heuristic or
numerical in nature. This left some room for scepticism and so
it is important that there is now a rigorous result by Dafermos
\cite{dafermos01} showing the development of weak null singularities
from regular initial data and the occurrence of the associated
effect of mass inflation.

That this kind of mathematical confirmation is not superfluous is
shown by the story of the supposed stability of Cauchy horizons
in the presence of a positive cosmological constant. This was
suggested in \cite{mellor90} and led to several papers coming to 
similar conclusions over a period of several years. However it turned 
out in the end \cite{brady98} that these
conclusions were incorrect. The original work had observed that
one mechanism leading to the instability of Cauchy horizons did
not seem to work in that context. It turns out, however, that a
different mechanism does appear to work. This throws some light on
the difference between heuristic arguments and rigorous ones.
In a mathematical argument all objections to a particular statement
must be effectively ruled out. This means that a mathematical 
approach to a given problem is often much harder than a heuristic 
one. Once, however, a mathematical argument has been successfully
completed it rules out the presence of any other effect that was
not thought of the first time around.

\section{Numerical relativity}

This section is concerned with the relation of the analytical
approach to the Einstein equations to numerical relativity.
We have already seen one kind of interaction, which is that
numerical results suggest a theorem which is later proved. 
Now the other direction will be considered, i.e. the question
whether analytical results can be of use to numerical relativists
in their choice of methods.

A lot of numerical relativity has been done using what are often
called the ADM equations. What is meant by this is the Einstein
evolution equations written in $3+1$ form in terms of the induced 
metric and second fundamental form of a foliation with a fixed choice
of lapse and shift. In fact these equations were in the literature
long before the works of Arnowitt, Deser and Misner (cf. \cite{foures48}).
Despite this the term \lq ADM equations\rq\ will be 
used in this now widespread sense in the following, whereby
the lapse and shift are restricted to take the simplest form
that they are identically one and identically zero respectively.
There is also an ambiguity due to the fact that the evolution
equations are only defined modulo the constraints. Here only 
the vacuum equations are considered and the condition that the
spatial components of the spacetime Ricci tensor vanish is taken
to define the evolution equations. In fact the definition of the
second fundamental form will be substituted into its evolution
equation to produce a second order equation for the metric of the
form $\d_t^2 g_{ab}=-2R_{ab}+\ldots$. Here only the terms containing
second derivatives of the metric have been written out.

Is the Cauchy problem for the ADM equations as just defined well-posed 
in the space of $C^\infty$ functions or in Sobolev spaces? Apparently
nobody knows. It is known that there is a smooth solution of the ADM
equations corresponding to any smooth initial datum satisfying the 
constraints and that the solutions depend continuously on the
initial data. This is proved indirectly, for instance by the use
of harmonic coordinates. This is all that is needed for the analytic
theory of the Einstein equations. In numerical relativity, however,
it cannot be expected that the initial data satisfy the constraints
{\it exactly}. Thus it is of interest to know whether the ADM
equations are well-posed as a system of evolution equations,
independent of whether the constraints are satisfied or not. It
has been shown by Choquet-Bruhat \cite{choquetunpub} that they are 
well-posed in Gevrey spaces, but the corresponding question for 
functions which are merely smooth, or belong to a Sobolev space, is open.

How can it be shown that a system of evolution equations is well-posed? 
This is related to the notion of hyperbolicity, which is at least
intuitively equivalent to well-posedness of the initial value problem
for appropriate initial data. It is
perhaps useful to start by looking at the notion of ellipticity.
Suppose that a nonlinear system of partial differential
equations is given. Its ellipticity can be determined as
follows. The system is elliptic if and only if its linearization
about any given solution is elliptic. With this rule it then
remains to define when a linear system is elliptic. To do
this take the principal part (the part containing the derivatives
of the highest order) and replace each derivative $\d_i$ by $\xi_i$.
This gives a homogeneous matrix-valued polynomial in $\xi$, called
the principal symbol. The system is called elliptic if the principal
symbol is invertible in the sense that for any $\xi\ne 0$ the values 
of the principal symbol are invertible matrices. More generally the
set of non-zero $\xi$ for which the symbol is not invertible is
called the characteristic set. Thus a system is elliptic if the
characteristic set of its linearization is always empty. 

Analogously to the elliptic case a non-linear system can be defined to
be hyperbolic if its linearization about any solution is hyperbolic.
Then it remains to define hyperbolicity in the linear case. There
are criteria for hyperbolicity in terms of the principal symbol, but
unfortunately, in contrast to the elliptic case, there is no simple
and general condition. The characteristics are given as the roots of
the characteristic polynomial, which is the determinant of the principal
symbol. In general, for a system of $k$ equations of order $n$ the
characteristic set has $kn$ sheets in the complex domain, counting
multiplicity. If a system is hyperbolic then there must be $kn$ real sheets.
In this case it is said that all characteristics are real. It was shown
by Friedrich \cite{friedrich96} that all the characteristics of the ADM
equations are real. This is a positive sign but unfortunately does not
suffice to show that the equations are hyperbolic.  

Note that in order to check that a system is hyperbolic it does not 
suffice to verify that the Cauchy problems for all one-dimensional 
reductions are well-posed. An example of this is given in \cite{courant}. 
This equation is
\be
u_{tttt}-2u_{ttxx}-3u_{ttyy}+u_{xxxx}+3u_{xxyy}+2u_{yyyy}+u_{xxy}=0
\ee
It has an ill-posed Cauchy problem. Now consider solutions of the
special form $u(t,x,y)=\tilde u(t,x\cos\alpha+y\sin\alpha)$. For each $\alpha$
this reduces the original equation to an equation for $\tilde u$ in
one space dimension. This reduced equation has a well-posed Cauchy
problem for each fixed value of $\alpha$. An important feature 
of this system is that the characteristics are of variable multiplicity.
This example concerns a single higher order equation but presumably
similar examples can be constructed for systems of first order equations.

To understand some of the issues involved in well-poseness it is useful 
to consider the trivial equation $u_{tt}=0$ for a function $u(t,x)$ of two
variables. With initial data $u(t,0)=f(0)$ and $u_t(t,0)=g(x)$ the
solution is $u(t,x)=f(x)+tg(x)$. If the equation were hyperbolic,
like the wave equation, it would be natural to take $f$ in the Sobolev
space $H^s$ and $g$ in $H^{s-1}$. Then the restriction of the solution 
to any other hypersurface of constant $t$ would be in $H^s$. However 
the explicit solution above is only in $H^{s-1}$ at later times.
The equation is hyperbolic, but in a weaker sense than the wave 
equation since it loses one derivative. What is worse is that 
the hyperbolicity can be destroyed by adding lower order terms.
The equation $u_{tt}=u_x$ (sideways heat equation) is not hyperbolic.
It is not possible to bound the $H^s$ norm at a given time $t_0>0$
in terms of Sobolev norms of the initial data. This can be seen by
doing a Fourier transform in $x$.

If the ADM equations are linearized about flat space the result
has some similarity to the equation $u_{tt}=0$ and is weakly
hyperbolic. If the full ADM equations are well-posed then
it is reasonable to expect that their linearization about any 
background will have the same property. Some examples of 
linearization about simple non-trivial backgrounds will now be 
considered. The equations bear some resemblance to the equation 
$u_{tt}=u_x$ above. This approach has so far led to no evidence
for ill-posedness of the ADM equations but it has not been carried
very far. Perhaps a different choice of background would lead to a 
different result. Some possible directions to pursue will now be
indicated.

In the examples to be considered here the metric can be written 
in the form
\be\label{gowdy}
ds^2=-dt^2+e^{2\alpha(t,x)}dx^2+e^{2\beta(t,x)}dy^2+e^{2\gamma(t,x)}dz^2
\ee 
Ideally we would like a background metric such that the linearized 
Einstein equations have constant coefficients. Then taking a
Fourier transform in space and time reduces the study of the
growth of linearized solutions in time to an algebraic problem.
It is natural to seek solutions with this property among spatially
homogeneous solutions. As a first example, consider the Milne
model which is given by (\cite{wainwright97a}, p. 194):
\be
ds^2=-dt^2+t^2(dx^2+e^{2x}(dy^2+dz^2))
\ee
This is of the form (\ref{gowdy}) with $\alpha=\log t$ and
$\beta=\gamma=\log t+x$. Linearizing about this solution and denoting
the linearized quantities corresponding to $\alpha$, $\beta$ and
$\gamma$ by $\hat\alpha$, $\hat\beta$ and $\hat\gamma$ respectively
leads to a system with coefficients depending explicitly on $t$.
However this dependence on $t$ is sufficiently simple that 
introducing a new time coordinate by $\tau=\log t$ reduces the
linearized Einstein equations to a system with constant coefficients.
A special case consistent with the equations is obtained by setting
$\beta=\gamma$. With $D=\d/\d\tau$ and ${}'=\d/\d x$ the system
becomes
\bea
&&-D^2\hat\alpha=3D\hat\alpha+2D\hat\beta+2\hat\alpha'+4\hat\alpha
-2\hat\beta''-4\hat\beta'     \\
&&-D^2\hat\beta=D\hat\alpha+4D\hat\beta+\hat\alpha'+4\hat\alpha
-\hat\beta''-4\hat\beta'
\eea
Note that while the equation for $\hat\beta$ contains the second 
spatial derivative of $\hat\beta$ the equation for $\hat\alpha$ does
not contain the second spatial derivative of $\hat\alpha$ but does
contain its first spatial derivative. This is a reason to suspect
that well-posedness might fail as a consequence of the model equation
discussed above. Now the equation will be Fourier transformed in 
space and time. Let $\omega$ and $\xi$ be Fourier variables 
corresponding to $\tau$ and $x$ respectively. Looking for solutions
of the form $e^{i(\omega \tau+\xi x)}$ leads to consideration of the
roots of the cubic equation for $\omega$ with coefficients 
depending on $\xi$
\be
\omega^3-7i\omega^2+(-\xi^2+2i\xi-14)\omega+(i\xi^2+2\xi+8i)=0
\ee
By inspection $\omega=i$ is a root. It then only remains to solve
a quadratic equation for $\omega$ to get the other two roots. These
given by $\omega=2i+\xi$ and $\omega=4i-\xi$. The question of interest for
well-posedness is whether the imaginary parts of the roots of the cubic 
equation for $\omega$ are bounded functions of $\xi$. In this example
this is the case and no evidence for ill-posedness is obtained.  

Consider next the Joseph solution, which is given by 
(\cite{wainwright97a}, p. 197):
\be
\sinh 2\tau [-d\tau^2+dx^2+A(\tau)e^{2x} dy^2+A^{-1}(\tau)e^{2x} dz^2]
\ee
where $A(\tau)=(\tanh t)^{\sqrt 3}$. Introduce a new time coordinate $t$
by means of the relation $dt=(\sinh 2\tau)^{1/2} d\tau$. Then the form
(\ref{gowdy}) can be obtained by setting $\sinh 2\tau=e^{2\alpha(t)}$,
$A(\tau)e^{2x}\sinh 2\tau=e^{2\beta(t,x)}$ and 
$A^{-1}(\tau)e^{2x}\sinh 2\tau=e^{2\gamma(t,x)}$. In this case the
coefficients of the linearized equations depend on $t$, although
the dependence is not too complicated. One way to try to get an idea
about well-posedness is to freeze the coefficients at a fixed time
$t$ and do a Fourier tranform of the resulting system with constant
coefficients. This has not been carried out. The conclusions to
be drawn from these examples is that there are many calculations 
which might be done which could lead to a better understanding of
the well-posedness or otherwise of the ADM equations. If the system
is ill-posed, one single example could be the basis for a proof of
ill-posedness of the full system. Unfortunately it not clear at
this time what is the best place to look for an example of this
kind.

Since a failure of well-posedness means a failure of continuous 
dependence of the solutions on the initial data and the initial
data in a numerical problem are never exact the significance of
results obtained using a numerical calculation with an ill-posed
system is quite unclear. On the other hand it is known that a
discretization of a partial differential equation may lead to
a scheme which approximates a different equation better than the
original one. (A well-known example of this is the introduction
of numerical viscosity leading to a good approximation to a
parabolic equation.) The ramifications of the well-posedness or
otherwise of the ADM equations are still rather unclear.

In numerical relativity it is not just the well-posedness of the pure
initial value problem which is important. It is necessary to impose
boundary conditions and this leads to new difficulties. Friedrich
and Nagy \cite{friedrich99a} have proved the well-posedness of certain 
initial boundary value problems for the (vacuum) Einstein equations but 
much more remains to be done. (See also \cite{szilagyi01} for some 
further recent work on the subject.)

An interesting puzzle on the interface between numerical and 
analytical work is provided by the BSSN system \cite{shibata95}, 
\cite{baumgarte99a}. It is apparently much better for numerical 
calculations than the ADM equations in some situations. It is not known 
to be hyperbolic although certain deformations of it were shown to give 
rise to symmetric hyperbolic systems in \cite{friedrich00} and 
\cite{frittelli99}. Different possible explanations of its superiority 
are discussed in \cite{alcubierre00a} and \cite{szilagyi99}   

\section{Outlook}

This section collects a few comments on some research directions which may
become important in the near future, without any claim to completeness.

A topic which can be expected to play a major role in the applications of
partial differential equations to general relativity in the future is that
of the global stability of important explicit solutions. Until recently the 
only results of this kind were the theorems of Friedrich \cite{friedrich86} 
on the stability of de Sitter space and Christodoulou and Klainerman 
\cite{christodoulou93} on the stability of Minkowski space. Now there 
are new developments in this area. Andersson and Moncrief have proved 
the stability of a vacuum cosmological model (the Milne model) in the 
expanding direction (see \cite{andersson99}) while Choquet-Bruhat and 
Moncrief \cite{choquet01}. 
have obtained a similar result for a vacuum cosmological model of Bianchi type 
III under the restriction that the perturbations are assumed to have one 
Killing vector. The Killing vector helps, but the Milne model is easier 
in other ways. It should be possible to generalize these results in many
ways (different backgrounds and matter models).

The results of \cite{andersson01} discussed in section 3 point the way to the 
possibility of proving the stability of the singularity in Friedmann
models with a scalar field. Combining this with the results on the
expanding phase mentioned above could lead to theorems on the global
stability of open Friedmann models with suitable matter content. The
stability of the singularity alone should allow conclusions about the 
stability of closed Friedmann models.  

A subject of capital interest is that of the stability of the Schwarzschild
and Kerr black holes. No nonlinear results are available at present although
there has been recent progress concerning linear stability \cite{beyer00},
\cite{friedman}. The nonlinear stability of the Kerr solution is one
of the outstanding open questions on the applications of partial 
differential equations to the Einstein equations.

The theme of critical phenomena has given rise to a considerable literature
in general relativity. All of it is based on numerical calculations.
The subject has been too tough to yield to analytical attack up to now.
A promising approach to changing this unsatisfactory state of affairs 
is to study model problems. In fact it turns out that critical behaviour
in gravitational collapse is not an isolated phenomenon within the field
of partial differential equations. It is more likely to be typical of a
kind of phenomenon which is widespread among solutions of nonlinear
evolution equations. For instance Bizo\'n, Chmaj and Tabor\cite{bizon00a}
have studied wave maps in two space dimensions while Bizo\'n and Tabor
\cite{bizon01a} have studied the spherically symmetric Yang-Mills equations 
in space dimensions greater than three. It has not yet been possible 
to handle the dynamics analytically, but at least the critical solution 
itself has been proved to exist in one case \cite{bizon99a}. 

The interaction between numerical and analytical work, some aspects of
which were described in section 4, has not really taken off, in the sense
that a lot of practical computations in numerical relativity are little
influenced by analytical developments. A notable exception is the numerical
implementation of the conformal field equations, which has recently
been reviewed by Frauendiener \cite{frauendiener00}. It will be interesting
to see whether this approach to numerical relativity eventually proves
superior to more pedestrian methods.

\section*{Acknowledgements}

I am grateful to Piotr Bizo\'n for discussions.

\end{document}